# Magnetism in Metastable and Annealed Compositionally Complex Alloys

Nan Tang,[1,†] Lizbeth Quigley,[1,†] Walker L. Boldman,[1,†] Cameron S. Jorgensen,[1] Rémi Koch,[2] Daniel O'Leary,[2] Hugh R. Medal,[2] Philip D. Rack,[1,3] Dustin A. Gilbert[1,4*]

[1] Department of Materials Science and Engineering, University of Tennessee, Knoxville, TN, 37996 USA
[2] Department of Industrial and Systems Engineering, University of Tennessee, Knoxville, TN, 37996 USA
[3] Center for Nanophase Materials Sciences, Oak Ridge National Laboratory, Oak Ridge, TN, 37831 USA
[4] Department of Physics and Astrophysics, University of Tennessee, Knoxville, TN, 37996 USA

[†] *Authors contributed equally to the work*

**Abstract:**

Compositionally complex materials (CCMs) present a potential paradigm shift in the design of magnetic materials. These alloys exhibit long-range structural order coupled with limited or no chemical order. As a result, extreme local environments exist with a large opposing magnetic energy term, which can manifest large changes in the magnetic behavior. In the current work, the magnetic properties of (Cr, Mn, Fe, Ni) alloys are presented. These materials were prepared by room-temperature combinatorial sputtering, resulting in a range of compositions with a single BCC structural phase and no chemical ordering. The combinatorial growth technique allows CCMs to be prepared outside of their thermodynamically stable phase, enabling the exploration of otherwise inaccessible order. The mixed ferromagnetic and antiferromagnetic interactions in these alloys causes frustrated magnetic behavior, which results in an extremely low coercivity (<1 mT), which increases rapidly at 50 K. At low temperatures, the coercivity achieves values of nearly 500 mT, which is comparable to some high-anisotropy magnetic materials. Commensurate with the divergent coercivity is an atypical drop in the temperature dependent magnetization. These effects are explained by a mixed magnetic phase model, consisting of ferro-, antiferro-, and frustrated magnetic regions, and are rationalized by simulations. A machine-learning algorithm is employed to visualize the parameter space and inform the development of subsequent compositions. Annealing the samples at 600 °C orders the sample, more-than doubling the Curie temperature and increasing the saturation magnetization by as much as 5×. Simultaneously, the large coercivities are suppressed, resulting in magnetic behavior that is largely temperature independent over a range of 350 K. The ability to transform from a hard magnet to a soft magnet over a narrow temperature range makes these materials promising for heat-assisted recording technologies.

## Introduction

Compositionally complex materials (CCMs) are alloys which are structurally ordered but leverage their chemical disorder to realize unique functional properties; while their structural analogues, termed high entropy alloys (HEAs), have been heavily investigated over the recent 15 years, their functional properties have not been as well explored.[1-3] The relatively few investigations that have been performed on CCMs, e.g. functional HEAs, report exciting, unexpected behavior, including non-traditional superconductivity,[4] promising thermoelectricity,[5,6] and ferromagnetism with extreme-coercivity[7,8] and magneto-structural coupling.[9] Large distributions in the underlying structure in these materials – including the electronegativity, bonding, atomic masses, and sizes – have the potential to alter the resultant physical properties.[7] While some of the properties manifest as an average of the collective behaviors, properties which are determined by localized interactions, such as magnetism, will result in a complex mixture of properties.

Most previous reports on CCMs have relied on preparing compositions which exhibit a single structural phase after long annealing treatments. This procedure is a vestige of structural HEAs, which rely on thermally induced diffusion to realize the chemical mixing in bulk synthesis, limiting the systems to the thermodynamically stable configurations. The work presented here uses room-temperature combinatorial sputtering to realize thin-film samples which are chemically intermixed without heating. By using sputtering, the atoms arrive on the substrate with some kinetic energy, allowing local surface diffusion to achieve ordering, but limited bulk diffusion.[10] The films thus have local structural order, but may not be in their thermodynamically stable phase, opening up a parameter space not accessible by traditional fabrication techniques.[11, 12] The wafers are then annealed, allowing us to measure the thermodynamically stable phase. Using a combinatorial approach, a range of compositions can be prepared on a single wafer.[13, 14] This work compares the effects of composition and ordering by measuring both an as-grown sample and an annealed sample.

The current work investigates magnetic ordering in a metallic alloy of (Cr, Mn, Fe, Ni). These elements were chosen due to their mixed magnetic interactions in single- and binary-component metallic alloys: Fe-Fe, Fe-Ni, and Ni-Ni are well known ferromagnetic (FM) formulations; Fe-Mn, and Ni-Mn, are antiferromagnetic (AFM); Mn-Mn is paramagnetic at any temperature. The Cr-Cr can form a spin-density wave due to nesting within the Fermi surface in the body centered cubic (BCC) structure, which is not expected to occur in the HEA lattice, due to the absence of a regular magnetic lattice and is treated as non-magnetic. Fe-Cr and Ni-Cr can have magnetic ordering on the Cr site for small at.% Cr (< 20 at.%[15] and 10 at.%,[16] respectively); we therefore also consider these interactions and Mn-Cr to be non-magnetic for this work. Previous works have performed theoretical investigations of these materials.[17-19] Presuming a BCC structure, with a central atom magnetically coupled to its 8 nearest neighbors, located at the vertices of the unit cell; there are 126 chemically unique unit cells. The dominant energy responsible for magnetic interactions is the exchange interaction, which is typically considered to dominate the spin alignment between two adjacent atoms (< 3 Å), but influences magnetic ordering over the exchange length, ≈50 Å.[20-22] In the presented alloys, it is virtually guaranteed that there will be FM, AFM and non-magnetic unit cells within the exchange length. As a result, it is unclear what the resultant magnetic ordering will be, but one might expect coupling between FM and AFM phases, resulting in exchange bias effects, including increased coercivity and loop shift. However, it is also well known that there is a volume scaling to the rigidity of AFM phases,[23-25] which contributes to an increased coercivity but no loop bias. In these systems, the antiferromagnet is 'dragged' by the reversing ferromagnet. Having a large coercivity but no bias results in a large energy product applicable to systems with symmetric bipolar fields – including spintronics, hard drives, and motors/generators. Another possibility is that the competition between the FM and AFM phases gives rise to frustrated ordering and eventually a spin-glass behavior. This material has also been presented as a promising structural material.[26]

This work shows that nanocrystalline as-grown CCMs present several surprising and exciting magnetic properties which are virtually absent in the ordered sample. The as-prepared alloys have a Currie temperature, $T_C$, ranging from room temperature (RT) to <50 K. Cooling below an apparent $T_C$ results in a square hysteresis loop with an extremely small coercivity (<1 mT), making it a very soft ferromagnet. However, at low temperatures, the coercivity, $H_C$, increases to >450 mT in some compositions, comparable to some high-anisotropy ferromagnets.[27] The swing between being a soft and hard magnet in a single phase material is highly uncommon. Annealing the samples increases $T_C$ to >400 K in virtually all samples and increases the saturation magnetization, $M_S$, by up-to 5×. However, the coercivity becomes constant at ≈45 mT for all temperatures and compositions. While ordering in metals is well known to change the magnetic properties, especially the magnetocrystalline anisotropy,[27] the simultaneous large changes in $M_S$, $T_C$ and $H_C$ are atypical of metals. Modeling suggests that the large change in coercivity can be attributed to the composite magnetic structure consisting of ferromagnetic clusters with decorating antiferromagnetic clusters. The measured samples were used to train a machine learning algorithm and predict the magnetic behavior and energy product of unmeasured and unrealized samples. The properties of these materials may have applications in heat assisted magnetic recording (HAMR) or high energy product permanent magnets.

**Results and Discussion**

*Film Structure*

Thin films of (Cr, Mn, Fe, Ni) alloys, 300 nm thick, were prepared via combinatorial sputtering on a 100 mm diameter (001)Si wafer coated with 85 nm of thermally grown $SiO_2$, resulting in a compositional gradient across the wafer. Using energy dispersive X-ray spectroscopy (EDX) the relative composition was measured at 17 points, shown in **Figure 1a-d**. X-ray diffraction of the as-grown wafer spanning across the equatorial belt of the wafer from (Cr-27%, Mn-43%, Fe-13%, Ni-17%) to (Cr-53%, Mn-7%, Fe-31%, Ni-9%), Figure 1e, shows a narrow peak attributable to the crystalline substrate, and broader peaks at 44.2° and 63.9° attributable to the film. The film peaks indicate the as-grown sample has a body-centered cubic (BCC) structure, with these peaks identified as the (110) and (200), giving a lattice parameter of 2.89 Å.[28,29] Sputter deposited metals with a BCC structure are well-known to exhibit (110) and (200) texturing, with Cr being a prototypical example.[30] The native structure of Cr and Fe are BCC, while the native structures of Ni and Mn are face centered cubic (FCC) and hexagonally close packed (HCP), respectively. The measured lattice parameter of 2.89 Å is similar to that of Fe (2.87 Å) and Cr (2.91 Å). A sister sample was grown and then annealed in vacuum at 973 K for 1 hour. This heat treatment is expected to achieve a higher quality of ordering in the thermodynamically stable phase at 973 K; the system is then cooled passively in vacuum over several hours. X-ray diffraction patterns of the annealed samples are shown in Figure 2a and identify four distinct phases: FCC (A1, $Fm\overline{3}m$), BCC (A2, $Im\overline{3}m$), σ-FeCr ($A_b$, $P4_2/mnm$),[31] and MnO (MnO, $Fm\overline{3}m$);[28,29] example patterns for each phase are shown in Figure 2b. Using the normalized peak areas, a rough phase fraction can be generated and the phases mapped onto the wafer, shown in Figure 2c. This map shows large regions of the sample, including the equacompositional point, which have a single structural phase (FCC and BCC), as-well as mixed phases with MnO precipitates and σ-FeCr. As expected, the peaks are much narrower, consistent with increased grain sizes and improved local structural ordering after annealing.

The microstructure of the films was measured using scanning electron microscopy (SEM), shown in the Supplementary Material Figure S1.[32] As-grown films show triangular shaped grains with a typical size of 300 nm² - 500 nm² with no apparent phase separation, consistent with the XRD. After annealing, the microstructure has spherical grains with a similar average size of 500 nm². Approaching the Mn source, small triangular grains precipitate, which we attribute to phase separated Mn clusters, oxidizing to the MnO observed in the XRD. Magnetic measurements were performed on a 50 nm thick film both as-deposited and after annealing – also at 973 K. After growth, the films were spin coated with a polymer photoresist for protection against oxidation and particulates, then diced into 5×5 mm² chips following a regular grid pattern. Magnetic measurements were performed on 18 chips from the as-grown wafer and 11 from the annealed wafer; the 11 chips from the annealed wafer were chosen as compositional counterparts from the as-grown wafer.

With the large number of samples (29) the data quickly becomes convoluted and it is challenging to identify trends. In order to better convey the results, the magnetic measurements from a single representative sample (Sample K, $S^K$), with a composition of (Cr-33%, Mn-12%, Fe-39%, Ni-16%), are presented. The trends for Sample K are discussed in detail, including Figure 3, then trends are shown for other compositions in the subsequent section, including Figure 4.

*Focus on Sample K*

Sample K is a representative sample with a composition of (Cr-33%, Mn-12%, Fe-39%, Ni-16%); X-ray diffraction shows a single FCC structural phase. Major hysteresis loops for Sample K at temperatures between 2.5 K and 400 K are shown for the as-prepared and annealed samples in Figure 3a, b and c, d, respectively. It is immediately apparent that the saturation magnetization for the annealed samples is much higher than the as-grown films, increasing by 460% at 2.5 K. The saturation magnetization at each temperature is tabulated and plotted in Figure 3e, emphasizing the increase in magnetization in the annealed sample.

Traditional M-T measurement on the as-grown film were performed in a 5 mT field, also shown in Figure 3e, and indicate a $T_C$ of ≈290 K. However, at a $T$>290 K the field response still shows a closed, s-shaped loop, indicating the presence of a minority magnetic phase with a higher $T_C$. There is also an atypical flattening of the M-T curve at ≈50 K; this flattening will become an inflection and drop in apparent magnetization in other compositions. The M-T curve from the annealed sample is also shown and indicates a $T_C$ >> 400 K. Fitting the M-T curve from the annealed sample using a mean-field approximation identifies $T_C$≈680 K, more than double the value for the as-grown sample.

Another apparent difference between the as-grown and annealed loops is the coercivity trends ($H_C$ trends), shown in Figure 3f. At the lowest temperatures, the as-grown film has a large coercivity (250 mT at 2.5 K), which rapidly decreases to 10 mT at 50 K, a rate of 5 mT K$^{-1}$. Further increasing the temperature, the hysteresis loop remains open and square up-to 200 K. At this point, the coercivity decreases to less-than 1 mT and retains a square loop shape. Finally, at $T_C$, the field response becomes sigmoidal, with a closed loop and a linear field response through zero. It is interesting to note that the low-temperature increase in coercivity is coincident with the flattening of the M-T curve. In comparison, the annealed sample possesses an extremely stable coercivity of 43 mT - 56 mT across all temperatures. The slight increase in $H_C$ at the lowest temperatures in the annealed sample can be attributed to the suppression of thermally activated reversal.

Complementary measurements of the resistance versus temperature (R-T) show a continuous downward trend, as is typical for a metal. The continuous trend in the R-T plot suggests that the magnetic changes observed are not the result of transitions in the electronic or structural ordering.

For comparison, major hysteresis loops and M-T plot for an as-grown and annealed Fe thin film is shown in Supplementary Materials Figure S2. These plots show a stark contrast, with comparably little difference between the as-grown and annealed samples, with similar values of $H_C$, $M_S$, and $T_C$ and their mutual dependence on temperature.

*Trends in Composition*

The key observations for $S^K$ are that the as-grown system (1) possess FM low-anisotropy to high-anisotropy transitions by cooling from 290K to 200 K to 50 K, which (2) is absent in the annealed sample. (3) Annealing the samples induces a significant increase (>5×) in the $M_S$ and (4) increase (>2×) in $T_C$. (5) The M-T curve shows a distinct flattening coincident with the increase in $H_C$. Similar properties and trends are observed in all of the other samples. This section follows these features and discusses their trend versus composition.

Two-dimensional heat maps of $M_S$ at 2.5 K are shown in Figure 4a and b, comparing the as-grown and annealed samples, respectively. Noting the scale in the legend, annealing the sample increases $M_S$ at every point on heat map, independent of the composition, and are consistent with the results for $S^K$. A reasonable expectation might be that $M_S$ is the lowest on the Mn-Cr edge, while the Ni-Fe edge, consisting of both ferromagnetic elements, would have the largest $M_S$. Indeed, in the as-grown sample, this is approximately correct, however, for the annealed sample, the $M_S$ trend is rotated ≈90°. In the annealed sample, the lowest $M_S$ is clearly in the Mn corner, and increases approaching the Fe corner. Although it is not clear from the heat map, every sample that was measured – annealed or as-grown – had a ferromagnetic component, with $M_S$ as small as 1 emu cm$^{-3}$. Thus, verifying theme (3) from above: the annealing process significantly increases the saturation magnetization.

The data presented in Figure 4a and b are captured at 2.5 K in a 1 T field, representing the saturation magnetization, however, Figure 3 showed that there is a strong and unusual temperature dependence to the magnetic behavior. Thus, while panels a and b represent the $M_S$ versus composition, panels c and d present the magnetization versus temperature for 11 compositions, as-grown and annealed, respectively. First, considering the as-grown samples, the compositions close to the Ni and Fe edge (A, J and K, identified in Figure 4e) have the largest magnetization, in agreement with the heat map. These M-T plots present a curve typical of a ferromagnetic system approaching $T_C$. Moving further into the wafer, towards the Mn-Cr edge, the M-T curves show a significant decrease in the apparent magnetization occurring at ≈50 K, and leveling out at ≈30 K. This drop is observed in Samples B, C, G, H and I, but can be suppressed in the M-T curve by cooling in a large field (>300 mT), shown in Supplemental Figure S3. Thus, the drop in the M-T curve is not an authentic reduction in the magnetization, but rather the emergence of an energy term which orients moments away from the applied magnetic field. Finally, moving deep into the wafer, Samples D, E, and F, shown in the inset, have an exceedingly small magnetization, with a $T_C$ of ≈160 K, and a drop in magnetization at ≈50 K. In comparison, none of the annealed samples have a drop in the magnetization. Thus, verifying theme (5) from above: M-T curves for the as-grown samples can have an apparent drop in the magnetization, which is absent in the annealed samples.

Another notable effect of annealing is the impact on $T_C$: all of the as-grown samples have a $T_C$ < 350 K, while most of the annealed samples have a $T_C$ > 350 K. For the as-grown wafer, samples with the highest fraction of ferrous elements have the highest $T_C$; increasing the Mn and Cr content, samples C, H, I, and B have a $T_C$ of ≈160 K; samples D, E, and F, which have the highest Mn and Cr content, also become ferromagnetic at 160 K. The ferromagnetism in these samples is confirmed by the measurement of an open hysteresis loop. Thus, verifying theme (4) from above: the annealing process significantly increases the Curie temperature.

The coercivity of the as-grown and annealed samples were measured at 2.5 K and is plotted in Figure 4e and f. Coercivity is an arcane metric in that it is influenced by several microscopic variables, including $M_S$,[33] defect density type and distribution,[34] microstructure,[35, 36] magnetic anisotropy and anisotropy distribution,[27, 37, 38] and magnetostatic interactions.[39] In this case, the magnetization of the film is expected to reverse by a domain growth mechanism. As-such, the coercivity will be determined by the ability to nucleate domains and propagate them through the film;[40] being in-plane measurements, the magnetostatic energies do not contribute. In the presented data, at 2.5 K, the coercivity in the as-grown film, panel e, drops rapidly approaching the Ni-Fe edge, but is >300 mT for a majority of the sample. These hysteresis loops predominantly have a square loop shape similar to Figure 3a. The largest coercivity observed was 478 mT in the as-grown Sample I (Cr-45%, Mn-13%, Fe-29%, Ni-13%), which is comparable to some alloys of $L1_0$-FePt$_X$;[27] this work showed FePtCu ternary alloys with 300 mT < H$_C$ < 1200 mT. The reported $H_C$ values for the as-grown film at low temperatures are much larger than typical ferromagnetic metals, which are <50 mT for a comparable 50 nm thick film, e.g. Fe in Figure S2.

In comparison, the $H_C$ heat map for the annealed sample shows a much smaller scale, ranging between 11 mT and 59 mT. The coercivity of the annealed samples has a broad maximum located around Samples B, H and K, centered at (Cr-37%, Mn-17%, Fe-30%, Ni-15%), achieving a value of 59 mT. Thus, verifying themes (1 and 2): the as-grown samples have an extremely small coercivity at higher temperatures and show a divergent coercivity at low temperatures; this is suppressed in the annealed sample. The maximal energy product ($BH_{Max}$) for the samples were determined at low temperatures and are shown in Supplemental Figure S4.

These results reiterate the features reported in $S^K$ while simultaneously emphasizing the dependence on both composition and temperature. Also important is that the equicompositional point, indicated on the heat maps by the large black circle, is not a 'special' point. While this is expected to be the point of maximal chemical entropy, it does not result in any particular features in the magnetic behavior.

*Saturation Magnetization and Curie Temperature*

In the as-grown film, one expectation is that $M_S$ should increase due to the number of valence electrons, following a Slater-Pauling trend.[41-43] Based on the composition, the valence is calculated and plotted as contour lines in Figure 4a and b, with the maximal value of 7.8 occurring on the NiFe edge, and the lowest (6.7) occurring at the Cr corner. Following a Slater-Pauling trend, one would expect $M_S$ to be minimal in the Cr corner and maximal at the NiFe edge. Indeed, while not precisely aligned, $M_S$ does generally increase from the Mn-Cr edge to the Ni-Fe edge. While a Slater-Pauling trend may contribute to the relative trend, the absolute scale of $M_S$ is entirely too small. Using the lattice parameter of the as-grown sample, the average magnetization for each constituent atom is calculated ($\mu_B$ per atom) and plotted in Figure 5a; the Slater-Pauling curve is also plotted. While the as-grown sample follows a roughly linear trend with valence electron count, the scale is ≈30× smaller than the predicted value.

After annealing, the contour lines showing the valence in Figure 4b are aligned effectively orthogonal to the expected trend. Furthermore, the linear trend in Figure 5a is lost and the reported magnetization values are still ≈8× too small. The large difference between the Slater-Pauling prediction and the measured value, even at extremely low temperatures, likely indicates that a large fraction of the sample is either disordered or antiferromagnetically ordered. The Slater-Pauling curve plots the average moment per atom, which cannot be captured for paramagnetic or antiferromagnetic regions through standard magnetometry. Measurement techniques such as neutron diffraction can identify the antiferromagnetic moment per atom, but is not possible in this case due to the absence of long-range magnetic periodicity (ordering and site-by-site moment).

The low $T_C$ of the as-grown film is expected based on the reduced exchange interaction from having non-magnetic neighbors and frustration resulting from a combination of FM and AFM interactions. Furthermore, the as-grown BCC structure has a low packing density, resulting in a reduced exchange interaction and $T_C$. Annealing the sample induces local ordering, including the higher density face centered cubic structure. The improved structural ordering into a higher density phase (FCC) is expected to increase the effective exchange interaction between FM neighbors, resulting in an increased $T_C$.

*The Drop in the M-T Curve and Exchange Bias*

Within the M-T plots of the as-grown sample, Figure 4c, are several samples which show a notable decrease in the magnetization at ≈55 K, followed by a leveling out below ≈30 K. The magnitude of the drop, measured from the M-T curves with a cooling field of 5 mT, is shown in Figure 5b. As noted above, this feature can be suppressed by cooling in large magnetic fields (>300 mT) and therefore does not correspond to an authentic reduction in the saturation moment, but rather a re-alignment. This realignment event incurs an effective cost to the Zeeman energy since the moments are now aligned against the magnetic field, which must be compensated by another energy term. The magnitude of the re-alignment energy can be estimated by calculating the Zeeman energy resulting from the suppression of the magnetization drop at the suppression field: E= $M_{Drop} \times H_{Sup}$. This energy is calculated to be 1.48 J cm$^{-3}$ for a sample with a composition of (16%-Fe, 18%-Ni, 37%-Mn, 28%-Cr). In comparison, the anisotropy energies of hexagonally close-packed Co and $L1_0$ FePt are 0.52 J cm$^{-3}$ and 8.8 J cm$^{-3}$, respectively.[44]

We expect that the origin of this energy is a change in the exchange interactions. Specifically, without a structural transition or a change in the electronic ordering, the magnetocrystalline anisotropy is expected to change very slowly. Similarly, the magnetostatic energy varies with $M_S$, which changes slowly far from $T_C$. This leaves the exchange energy as the likely culprit. The change in the exchange interaction may be the result of an emergent antiferromagnetic ordering. The onset of antiferromagnetic ordering would be accompanied by corresponding exchange bias, which is observed and tabulated in Figure 5c at 2.5 K after field cooling in 1 T. The presence of exchange bias is strong evidence that emergent antiferromagnetism is the origin of the low-temperature anisotropy. However, antiferromagnetic ordering would not be suppressed by cooling in 300 mT, suggesting that the drop in the M-T curves is likely proximity regions which become demagnetized once the AFM ordering is established. Comparing the exchange bias to the drop in the M-T plot, Figures 4b and 4c, shows some correlation, however, the strongest exchange bias occurs at a higher Mn concentration. Similar to the coercivity, the magnitude of the bias is not a simple indicator of AFM strength or AFM-FM coupling, but comparing the energy of the exchange bias ($\approx M_S \times H_B$) provides a better insight, and is provided in Supplemental Figure S5. Another factor to consider is the increased coercivity at $T < 30$ K. Specifically, the Figure 3f shows an increase in coercivity in the as-grown films which increases below 30 K, coinciding with the leveling off in the M-T plots. This coincidence is emphasized in Supplemental Figure S6, which plots $H_C$-T and M-T for Sample C. Consistent with the above explanation, below 55 K an antiferromagnetic phase grows in volume and strength. As the AFM phase increases in volume, its effective anisotropy increases as-well,[24] resulting in the exchange bias effects. A low-anisotropy AFM can be 'dragged' by the reversing ferromagnet, resulting in an enhanced coercivity without exchange bias or asymmetry in the loop shape.[23, 45]

To explain these various magnetic observations, we present a physical model derived from the statistical distributions of the constituent atoms, shown illustratively in Figure 5d. This model treats the as-grown film as a site-by-site weighted random distribution of Fe, Ni, Mn, and Cr, and then determines the local FM and AFM interactions, shown as red and blue, and non-magnetic or frustrated regions, in white. Simulations are performed in the next section based off of this model, however, a physical description is provided here as it is relevant to the drop and the exchange bias. This model predicts small clusters of $Fe_XNi_Y$ which possess local ferromagnetic ordering. Even in a homogeneous material, the local atomic distribution will be determined by statistics, resulting in localized FM and AFM ordering. For higher Mn and Cr concentrations, these clusters are isolated from each other, potentially resulting in larger thermodynamic affects, analogous to superparamagnetism.[46] However, the larger effect is likely that the Mn and Cr inclusions behave as nucleation sites, precipitating low-field magnetic reversal. Thus, this model predicts a narrow hysteresis loop at high temperatures, which open at lower temperatures, precisely as observed in Figure 3. These clusters exert a proximity ferromagnetic interaction on their neighbors within the exchange length (typically considered $\approx 50$ Å).[20-22] This interaction can induce a magnetic polarization in these neighbors – especially in the frustrated regions.

Reducing the temperature, we propose that clusters of FeMn and NiMn establish AFM ordering. This ordering also exerts a proximity interaction which acts against the FM clusters, reducing any induced polarization and resulting in a glassy behavior. Also, in the boundary region between the AFM and FM clusters, the random distribution of interactions results in a high density of atoms which may be pinned to the antiferromagnet, relative to a traditional bilayer exchange biased system. These moments are analogous to the uncompensated spins at the interface of traditional exchange bias systems[47, 48] and exert a strong influence on the FM clusters. Coupling between the AFM and FM through these uncompensated moments will increase the effective anisotropy of the clusters, increasing the coercivity. Exchange bias behavior manifests four characteristic features: a horizontal and vertical loop shift, enhanced coercivity and loop asymmetry. Indeed, the measured loops which show loop shift also show the other three qualities; Sample C, located at a local maxima of exchange bias, has a vertical shift of +6%, a horizontal (field) shift of -110 mT, an enhanced coercivity discussed extensively above, and a loop asymmetry.

This model predicts three trends directly observed in the data. First, for the low Mn-concentration regions, the AFM clusters are small and isolated, while the FM regions are large and extended. It is well known that AFMs have a size-dependence to their anisotropy energy, and for small/thin AFMs, their orientation is 'dragged' by proximity interactions with a FM. This results in the absence of exchange bias but a persistent enhancement to the coercivity.[23, 24] Indeed, approaching the Fe-Ni edge, where the Mn content is significantly reduced, the coercivity does diverge at low temperatures, but there is no exchange bias.

The second prediction is that at high Mn concentrations, the AFM clusters become larger and extended, establishing a robust ordering with sufficient anisotropy to establish exchange bias. These clusters are comprised of FeMn and NiMn, which have Néel temperatures, $T_N$, of 490 K and 600 K, respectively.[22, 49, 50] The large extended AFM phase with high $T_N$ components establishes ordering at a much higher temperature and cannot be 'dragged'. Indeed, this is also observed in the data, as the samples near the Mn corner have an established exchange bias and much smaller $M_{Drop}$.

Interestingly, the phase with the largest coercivity is nearly 50% Cr and only 13% Mn, thus a vast majority of the exchange interactions will be non-magnetic or ferromagnetic. While Cr, in low concentrations with Fe and Ni, can possess AFM ordering, the fractions here are much higher. One possibility is that the Cr is participating in the AFM ordering. Another possibility is that the Cr is facilitating an RKKY interaction, which encourages an oscillating FM/AFM ordering, further frustrating the magnetism. The Fe-Cr system is well known to be one of the strongest RKKY systems.[51]

The third prediction is that the drop in the M-T plot can be suppressed by an applied magnetic field. Specifically, since the FM clusters align with the magnetic field through the Zeeman energy, the field encourages alignment of the interstitial frustrated regions with the FM particles. This manifests in the measurements as the suppression of the drop at large fields as seen in Supplemental Figure S3. The ability to suppress this drop also excludes simple antiferromagnetism as the origin since the spin-flop field is typically much larger than the 300 mT reported here.

None of the annealed samples show exchange bias (loop shift or an enhancement of the coercivity) at any temperature. These qualities may arise from four underlying mechanisms: (1) the annealed sample no-longer has antiferromagnetic ordering, (2) the AFM and FM phases are well separated, (3) the AFM has no net orientation and (4) the AFM clusters are too small to possess a robust anisotropy. Mechanism (1) is supported by the increased $T_C$, which indicates enhanced ferromagnetic exchange interactions. Furthermore, approximately half of the sample has BCC ordering, while the AFM ordering in FeMn and NiMn are in the FCC phase. Mechanism (2) is supported by the X-ray results, which indicate that, in some compositions, annealing causes phase separation, which may separate the FM and AFM phases. Mechanism (3) is likely, since the Néel temperature of FeMn and NiMn are 480 °C and 800 °C, respectively. As the sample is annealed there is no magnetic field to set the orientation, and so the AFM phases lack a long-range correlation, suppressing any net loop shift. We have no evidence addressing mechanism (3), however, both (3) and (4) would result in an atypically large coercivity; the annealed samples show a coercivity with is atypically large (≈50 mT) for a magnetic thin-film, which may be an indication of these phenomena. Alternatively, the increased coercivity could be the result of domain wall pinning by small AFM clusters, consistent with Mechanism (3). Thus, the absence of a loop shift or apparent Néel temperature does not preclude antiferromagnetism in these samples, and the magnetometry measurements support this possibility.

An alternative theory for the increased $T_C$, $M_S$ and $H_C$ in the annealed samples is that some high-anisotropy phases order during the thermal processing. Previous works have reported an emergent $L1_0$ phase within these compositions. The $L1_0$ phase of e.g. FeNi is known to be a high-anisotropy ferromagnet. Precipitating these phases within the film would act as magnetically hard pinning sites which pin the domain wall, resulting in an increased $H_C$. The $L1_0$ phase of FeNi has presented a challenge to prepare in its single phase,[52-54] high-anisotropy composition; HEAs/CCMs may present an opportunity to promote and stabilize this phase. While this phase has been reported previously,[28, 29] the indicative (100) peak in the XRD[27] and the splitting of the (200)/(002) peaks was not observed in any of the samples.

*Simulations of the Random System*

Simulations of the magnetic configurations and analytical calculations were performed using a simplified nearest-neighbor model. In these models, each lattice site was randomly identified as one of the constituent atoms by their compositional weight. The nearest neighbor interactions were tabulated, and the resultant cluster identified as nonmagnetic, ferro-, or antiferromagnetic, as described in the methods section. These computational results simulate the disordered systems and provide insight to the local magnetic ordering which experimental measurements only capture as a collective. The FeNiMnCr alloy was modeled as a weighted random distribution of 10,000 atoms on a 2D square grid for Sample A, $S^A$, which is approximately equicompositional, Sample C, $S^C$, which has the largest drop in the M-T plot, and Sample E, $S^E$, which has the exceedingly small $M_S$. The use of a 2D grid is a simplified system analogous to the classical XY model,[55] but captures the physical results surprisingly well. As described in Methods, the interaction for each cell relative to its nearest neighbors is calculated and averaged. The resultant magnetic structure is shown as an example in Figure 5d-f, and for each of the samples in Supplemental Material Figure S7, with FM regions shown in red, AFM regions in blue, and zero net interaction in white. The average cluster size and coverage (%) for the three highlighted samples are shown in **Table 1**.

|  | **Sample A ($S^A$)** (25:25:25:25) | | **Sample C ($S^C$)** (18:18:29:35) | | **Sample E ($S^E$)** (16:13:42:29) | |
|---|---|---|---|---|---|---|
|  | Size (cells) | Coverage (%) | Size (cells) | Coverage (%) | Size (cells) | Coverage (%) |
| *Non-Magnetic\** |  | 26% |  | 35% |  | 48% |
| *Balanced FM+AFM\** |  | 9% |  | 8% |  | 6% |
| *Total Magnetic\** |  | 64% |  | 57% |  | 45% |
| *Ferromagnetic* | 21 | 32% | 6 | 12% | 6 | 6% |
| *Antiferromangetic* | 10 | 20% | 10 | 25% | 9 | 14% |
| *Monochrome* | 319 | 68% | 32 | 44% | 14 | 27% |
| *Proximity Small Clusters* | ≤ 3 | 16% | ≤ 3 | 7% | ≤ 3 | 7% |
| *Isolated Small Clusters* | ≤ 3 | -4% | ≤ 3 | 13% | ≤ 3 | 18% |

*Analytical Calculation

***Table 1*** Calculated magnetic and nonmagnetic phases. Rows 1-3 are analytical calculations; Rows 4-6 are directly measured in the simulation; Row 7 is the difference between Row 6 and Rows 4 and 5; Row 8 is the difference between Row 3 and Row 6; cell 8 is negative as a consequence of subtracting the analytically calculated statistical value from a finite numerical simulation.. The "Size" column is the typical cluster size in cells, which is representative of lattice sites.

The calculated results agree well with the proposed mechanistic model and are consistent with the experimental results. Specifically, between $S^A$ and $S^C$, the FM cluster size is significantly reduced, which results in a reduced local exchange interaction which suppresses $T_C$, exactly as observed in the experiments. Furthermore, the FM cluster size does not change between $S^C$ and $S^E$, resulting in the same $T_C$, consistent with the experiments. Similarly, the reduced coverage of the FM clusters corresponds to a reduced $M_S$, related by a 3/2 power (due to scaling into the three-dimensional experimental parameter space). Simulations predict an $M_S(S^A)/M_S(S^C)$ ratio of 4.4 while experimentally it is 5.2. These results suggest that, despite this being grossly complex system, this simplistic model captures some of the key magnetic features.

These models also show that large fractions of the sample have no net magnetic interaction due to either the central cell being non-magnetic Cr, local balancing of the interactions, or being surrounded by non-magnetic interactions. The fraction of cells with these conditions can be calculated analytically. Subtracting these phases from the total simulation volume, the total expected fraction which is magnetically ordered can be calculated and is found to be significantly larger than the simulation results, e.g. FM+AFM. This difference is due to the exclusion of small clusters (≤ 3 cells) in the counting scheme. The distributions were re-analyzed by converting the image to a monochrome – removing the distinction between FM and AFM regions. In this analysis, small magnetic clusters adjacent to or embedded within larger clusters now contribute to the total magnetic ordering, while isolated small clusters remain uncounted. Using these values, the fraction of isolated and proximity small clusters can be calculated. The results show that, in $S^A$, the magnetic cluster size is 10× larger than the individual analysis, indicating that the FM and AFM clusters are highly intermixed, with the AFM and FM phases in direct contact. Direct contact between the AFM and FM phases results in exchange bias and proximity induced magnetic ordering.[21, 22, 47] Indeed, Figure 5c shows that $S^A$ is exchange biased, but it is not the largest bias. This is likely the result of the larger $M_S$ in $S^A$, which results in a larger Zeeman energy that competes with the exchange bias; the Zeeman cost to the exchange bias is shown in Supplementary Material Figure S5 and agrees with this interpretation. The analysis also reveals that the there is significant intermixing of the clusters, leaving very few isolated small clusters.

In addition to the large intermixing of FM and AFM phases in $S^A$, the FM phase along is also highly interconnected, forming a weakly coupled network. In effect, ferromagnetic filaments connect larger magnetic clusters. The XY model analogue of this system[55] and subsequent works,[56, 57] showed that these features can form 360° domain walls which possess an inherent topology. In this system, the combination of ferro- and antiferromagnetic components may be resolved by the formation of a short-range 360° domain wall.[57] Alternatively, FM clusters, connected through these filaments, would be highly-likely to trap pairs of 180° domain walls with a common chirality, thus forming a 360° domain wall. Such a feature is typically challenging to realize due to the large exchange energy, but in this system, the frustrated local magnetization may establish these features as a low-energy state. In contrast, in $S^C$ and $S^E$ there is a large fraction (13% in $S^C$, 18% in $S^E$) of the samples that are small isolated clusters. These small isolated clusters, defined as 3 or less cells, and the 'non-magnetic' clusters (8% in $S^C$, 6% in $S^E$) which result from balanced local interactions, are not expected to contribute to the magnetization. The proximity of small clusters and 'balanced' clusters may experience induced magnetic ordering from the larger clusters, which is overtaken below the AFM ordering or by large magnetic fields, resulting in the observed magnetic behavior. In contrast, the highly intermixed nature of $S^A$ likely suppresses these effects due to the larger cluster size, extended magnetically ordered phase, and contact boundary.

*Machine Learning Predictions*

One of the key challenges in the development of CCMs is navigating the vast parameter space provided by these multi-component alloys. The presented work measured 30 samples, consuming significant time and effort, yet offers only a small window into these alloys. Machine-learning (ML) algorithms deployed to predict the properties of unmeasured compositions and develop a map of the composition/property parameter space.[58] This map can be used to identify trajectories for subsequent samples and to identify trends, lending insight into the underlying physics.

Data from the as-gown samples was used to train two ML algorithms, Gaussian process regression (GPR) and random forest regression (RFR), which then generated predictions for the saturation magnetization, coercivity, and bias field. Unlike the interpolated data shown in Figures 1, 3 and 4, the ML algorithms can predict the properties of compositions which are not on the wafer and were never synthesized. This data can then be used to direct subsequent sample fabrication and optimize material properties. The saturation magnetization can be modeled using the 18 as-grown data points, plus additional points derived from the single-element and binary alloys on the boundary. A ML-derived contour plot for $M_S$ is shown in Figure 6a and reflects the general trends identified in Fig. 4a and b. The $M_S$ increases towards the Fe corner, and decreases towards the Mn corner (uncertainties shown in Fig. S8). While $M_S$ is relatively predictable, the coercivity and bias field depend strongly on the nanoscale distribution of the elements and magnetocrystalline anisotropy, intermixing of the magnetic phases, the film microstructure, and defect density. The ML derived contour plots for $H_C$ and $H_B$ are plotted in Fig 5b and c, respectively. Regression plots, comparing the experimental and predicted values are shown in Figure S8d-i, and regression metrics (Table S1) generally confirm the early suggestion: the $M_S$ regression shows good agreement, while the $H_C$ predictions are much less accurate. While the coercivity is expected to increase with decreasing $M_S$ due to the reduced influence of the magnetic field, the trend in $H_C$ shown in Fig 5b does not follow the $M_S$ trend. As noted above, the increased $H_C$ is the result of exchange bias between the AFM and FM regions. The exchange bias can be seen in the $H_B$ trend, Figure 6c, and is shown to increase towards the Mn boundary. Both $H_B$ and $M_S$ trends do not follow the $H_C$ trend, indicating that other mechanisms may be dominant, including potentially the pinning potential from the increased presence of non-magnetic Cr.

To improve the accuracy of off-wafer predictions we added known points in the configuration space (e.g., $M_S$ of Fe is known to be 1714 emu/cm$^3$). However, we found that adding these points reduced the accuracy of on-sample predictions (see regression plots in Figure S8d-i as well metrics in Table S1). In fact, including a large number of known points caused the R-squared value to become negative, indicating that the predictions are worse than using the average value as a prediction. For $M_S$ and $H_C$ the random forest regression predictions were more accurate, likely because regression trees are better able to model discontinuities. While the $H_B$ predictions were poor for both GPR and RFR, GPR showed a small amount predictive power, while the RFR predictions were worse than using the average value.

**Conclusion**

In summary, this work reports on the magnetic behavior of (Cr, Mn, Fe, Ni) CCMs, both as-prepared via sputtering and after ex-situ annealing. In the as-grown BCC-ordered sample, the samples have an exceedingly small coercivity (<1 mT), which increases slowly with cooling until T=50 K. At 50 K, an apparent drop is reported in the M-T plots and commensurate increase in coercivity, up-to 480 mT. Increasing the ferromagnetic fraction of the composition appeared to suppress the drop, but does not suppress the coercivity enhancement. In contrast, the annealed samples did not show a drop in M-T or divergent coercivity, but did possess a >5× increase in $M_S$ and >2× increase in $T_C$ compared to the as-grown sample. These results were compared against a simplified nearest-neighbor model, which showed excellent qualitative agreement with the experimental results. After training a machine-learning algorithm with these results, a map was developed to target the next iteration of samples. Together, these results identify interesting magnetic behaviors in metastable CCMs which are distinctly different from their annealed counterparts. This detail is critical since virtually all CCMs and HEAs studied previously have relied on high-temperature processing, leaving the former class of materials effectively unexplored. Finally, this work demonstrated that the equacompositional point was not particularly unique, as is often emphasized in CCMs/HEAs.

**Methods**

*Sample Fabrication*

Thin films (50 nm and 300 nm) of (Cr, Mn, Fe, Ni) were grown on thermally oxidized 100 mm Si wafers from elemental targets using confocal radio frequency sputtering. Deposition was performed at room temperature in a 5 mTorr Ar atmosphere. By not rotating the sample during deposition, a range of alloy compositions are prepared with the compositional ratios varying with position on the wafer. The composition was determined at 17 points using energy dispersive X-ray spectroscopy (EDX), then the composition across the wafer was determined using a 2D surface interpolation. The extremum corners of the sample were measured to be (Cr-68%, Mn-7%, Fe-17%, Ni-8%), (Cr-43%, Mn-36%, Fe-10%, Ni-11%), (Cr-24%, Mn-29%, Fe-23%, Ni-24%), and (Cr-32%, Mn-10%, Fe-44%, Ni-14%), correlating to the Cr, Mn, Ni and Fe corners of the wafer. By depositing at room temperature, the incident atoms do not have sufficient energy to diffuse on the surface, and thus develop a nanocrystalline structure. The microstructure of the 300 nm thick as-deposited film was measured with scanning electron microscopy (SEM). The SEM results showed smooth grains distributed uniformly across the surface, with an average area of 545 nm$^2$; using a spherical grain approximation, the grains have an average radius of 13 nm. X-ray diffraction (XRD) was performed on the 300 nm thick sample using Cu-kα X-rays (λ = 1.54 Å). A second 300 nm thick film was deposited and annealed at 973 K in high vacuum (P < 10$^{-7}$ Torr) for 1 hour. X-ray diffraction showed that the annealed sample possessed several narrow peaks, indicating the film had developed significant, long-range structural order.

A second set of 50 nm thick samples were grown and one was annealed after deposition. The 50 nm films were coated with photoresist and diced into 5×5 mm$^2$ chips for subsequent magnetic measurements.

*Magnetometry*

Magnetometry measurements were performed using a vibrating sample magnetometer (VSM) with an in-plane magnetic field, at 2.5 K, 10 K – 50 K in 10 K increments and 50 K – 400 K in 50 K increments. Magnetization versus field (M-H) hysteresis loops were measured by field cooling in 1 T from 400 K to 2.5 K then measuring with increasing temperature. The saturation magnetization, M$_S$, was determined by the magnetization at 2.5 K and 1 T; the coercivity and bias are calculated by the average and half-difference of the points H(M=0). Magnetization versus temperature (M-T) measurements were performed by heating the system to 400 K, applying a 5 mT and cooling to 2.5 K. Resistance measurements were performed following the same procedure.

*Simulations*

The FeNiMnCr alloy was modeled as 10,000 atoms on a 2D square grid. Each cell within the grid was filled by a random number generator with a value between 0 and 1; the elemental occupancy for each cell was determined by setting ranges defined by the composition. For example, cells with a value between 0-0.25 were defined as Cr, 0.25-0.5 as Mn, 0.5-0.75 as Ni and 0.75-1 as Fe, corresponding to a 1:1:1:1 ratio. The average magnetic interaction at each cell was calculated by summing the total number of FM and AFM neighbors, using the premise that Fe-Fe, Fe-Ni, and Ni-Ni are FM, Fe-Mn, Ni-Mn are AFM and Mn-Mn and anything with Cr are non-magnetic. This value was used to define a scheme, with red representing FM, blue as AFM, and white as non-magnetic. The color channels were split and analyzed using the particle analysis function in ImageJ. Conditions for the analysis enforced that at least 4 FM or AFM cells must be adjacent in order to count towards a resultant ordering; considering these values are the result of a sum of nearest neighbors, this 'cluster' emphasizes the influence of 12-14 atoms, depending on the cluster geometry. The image was converted to monochrome to determine the total fraction that possesses magnetic ordering. Analyzing the image as a monochrome captures the FM and AFM clusters and the clusters that are a mixture of FM and AFM ordering. Small clusters excluded in the original analysis but in proximity to other clusters would be counted in the monochrome analysis, but isolated clusters would not.

*Machine Learning Predictions*

The noise-free Gaussian process regression (GPR) approach used models a function as a stochastic process, using a Matérn kernel function to describe the correction between the function values at different points. In this work a kernel function was used that assumes that (Cr, Mn, Fe, Ni) configurations that are closer in parameter space will have measurements (i.e., saturation magnetism, coercivity, and bias field) that are similar, on average. Future work could relax this assumption by using a treed GPR model[59] that uses a data-driven approach to partition the parameter space, making it better suited to model discontinuities due to changes in phase. To implement the GPR model, a standard implementation of a single-task exact Gaussian Process regression model from the BoTorch package[60] was used. The random forest regression (RFR) uses the average of random regression trees (generated from samples of the data) as predictions. We used 1000 regression tree estimators in our implementation.

In addition to the samples taken from the thin film, the following theoretical values of saturation magnetism were added to the training data in order to improve the accuracy of the predictions (all values in emu cm$^{-3}$): $1714x + 484(1-x)$ for Fe$_x$Ni$_{1-x}$ for $x = 0,5,10,\ldots,95,100$; 0 for 100% Mn, 100% Cr; and $1714x$ for Fe$_x$Cr$_{1-x}$ for $x = 5,10,\ldots,95$. The bias field was assumed to zero for all of the aforementioned compositions. For 100% Fe the coercivity of a reference sample was measured as 46 mT. To aid in hyperparameter tuning, the training data inputs were normalized, and the training data outputs were standardized to have zero mean and unit variance. To improve model fit a sigmoidal transformation was used for the saturation magnetism data, and a log transformation was used for the bias field data. The predicted outputs were un-standardized and un-transformed prior to plotting.

## Data availability

The data that supports this work and the findings of this study are available from the corresponding author upon request.

**Acknowledgements**

P.D.R acknowledges support from the U.S. Department of Energy (DOE) under Grant No. DE-SC0002136. D.A.G., N.T., C.S.J. and L.Q. acknowledge support from the U.S. Department of Energy, Office of Basic Research CAREER program under Award Number DE-SC0021344.

**Author Contributions**

L.Q., N.T. and W.L.B. contributed equally to this work. Sample fabrication, X-ray diffraction, and microscopy was performed by W.L.B. and P.D.R. Magnetometry and transport measurements were performed by N.T., C. J., L.Q., and D.A.G. Data analysis was performed by L.Q. and D.A.G. Machine learning work was performed by R.K., D.O'L. and H.R.M. Experiments were designed by D.A.G., H.R.M. and P.D.R. Modeling was performed by D.A.G. The first draft of the text was written by L.Q., N.T. and D.A.G. All authors made contributions to and approved the final text.

**Corresponding Authors**

Dustin A. Gilbert: dagilbert@utk.edu

**Figures**

*Figure 1*

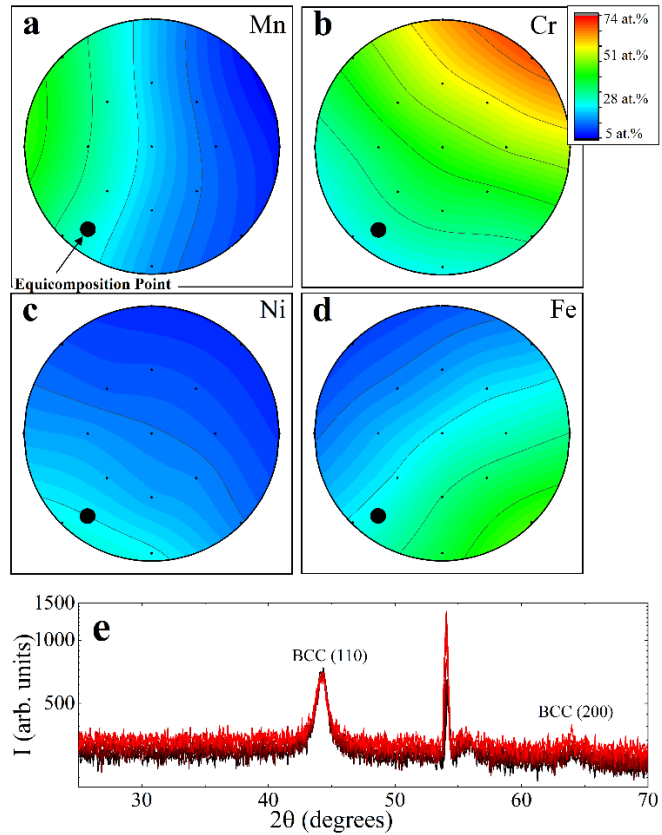

*Figure 1 Composition map showing at.% of (a) Mn, (b) Cr, (c) Ni, and (d) Fe. The circular illustration is representative of the 100 mm diameter wafer. (e) X-ray diffraction of the as-grown wafer measured across the equatorial band of the wafer (5 measurements), ranging between (Cr-27%, Mn-43%, Fe-13%, Ni-17%) in red and (Cr-53%, Mn-7%, Fe-31%, Ni-9%) in black.*

*Figure 2*

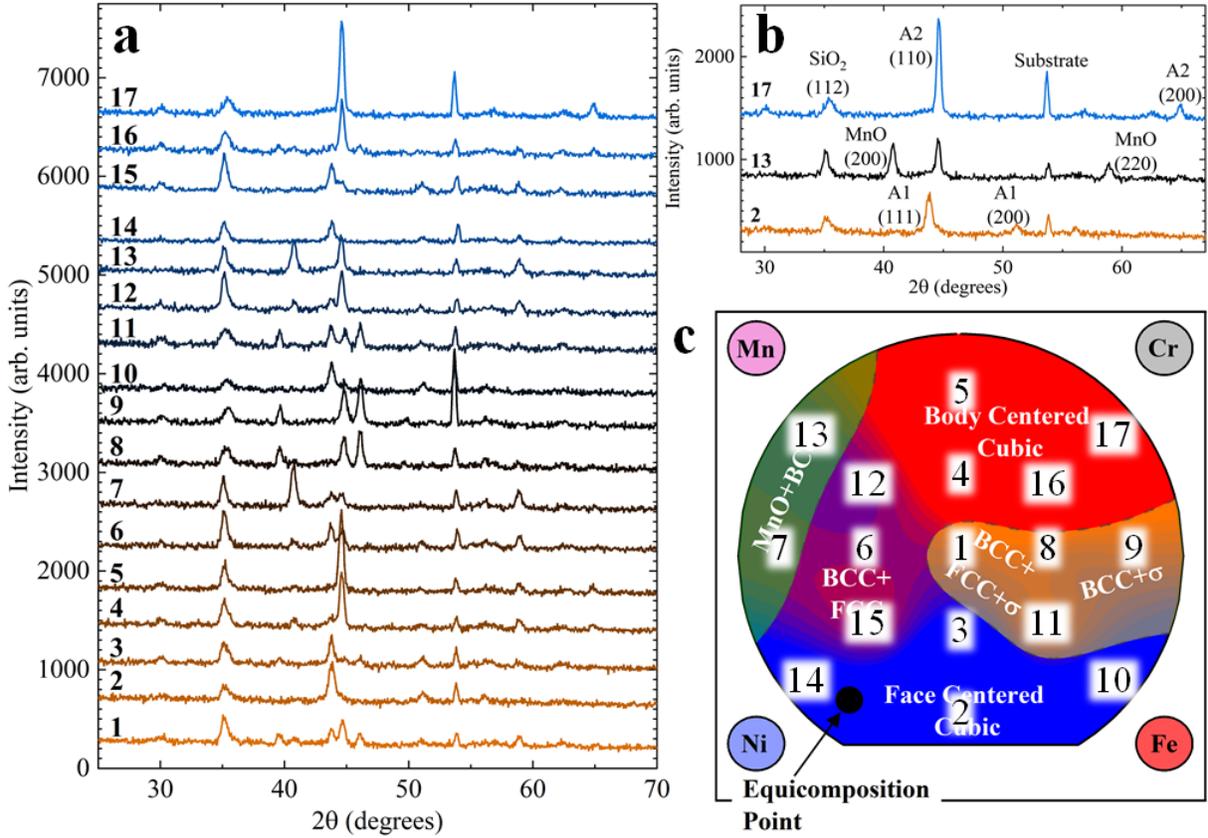

**Figure 2** *(a) X-ray diffraction data for the 17 measured points, (b) select measurements with identified peaks, and (c) map of phases on wafer, with measurement locations identified.*

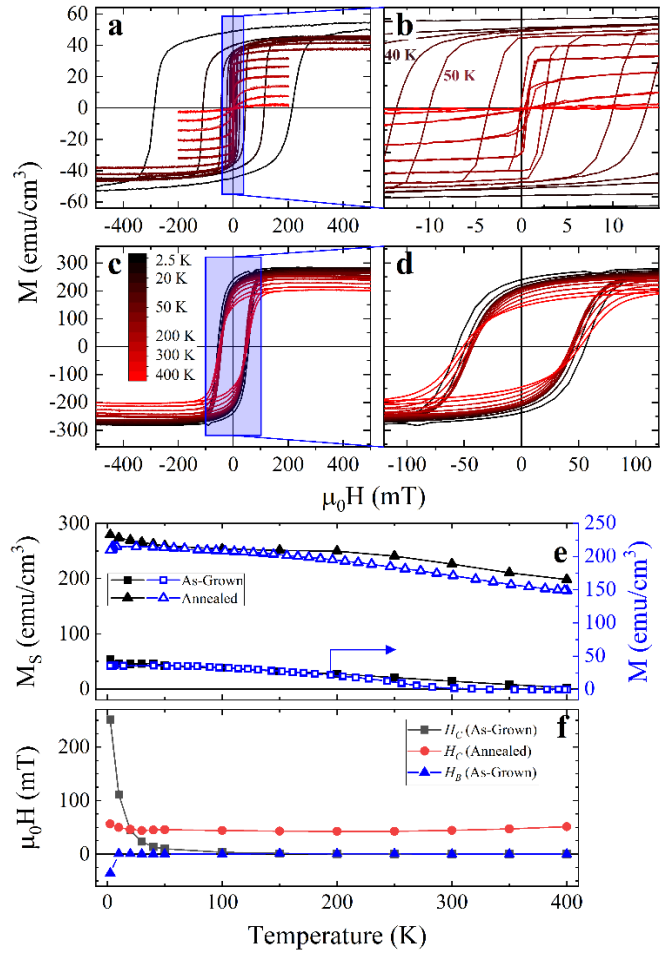

***Figure 3*** *Major hysteresis loops measured at 2.5 K – 400 K for $S^K$, (Cr-33%, Mn-12%, Fe-39%, Ni-16%) (a,b) as-grown and (c,d) after annealing. Panels on the right are zoomed in views of the shaded blue boxes. (e) Saturation magnetization (left, black, solid) and field cooling magnetization (right, blue, open) for the as-grown and annealed samples. (f) Coercivity of the as-grown and annealed samples and exchange bias of the as-grown sample versus temperature.*

*Figure 4*

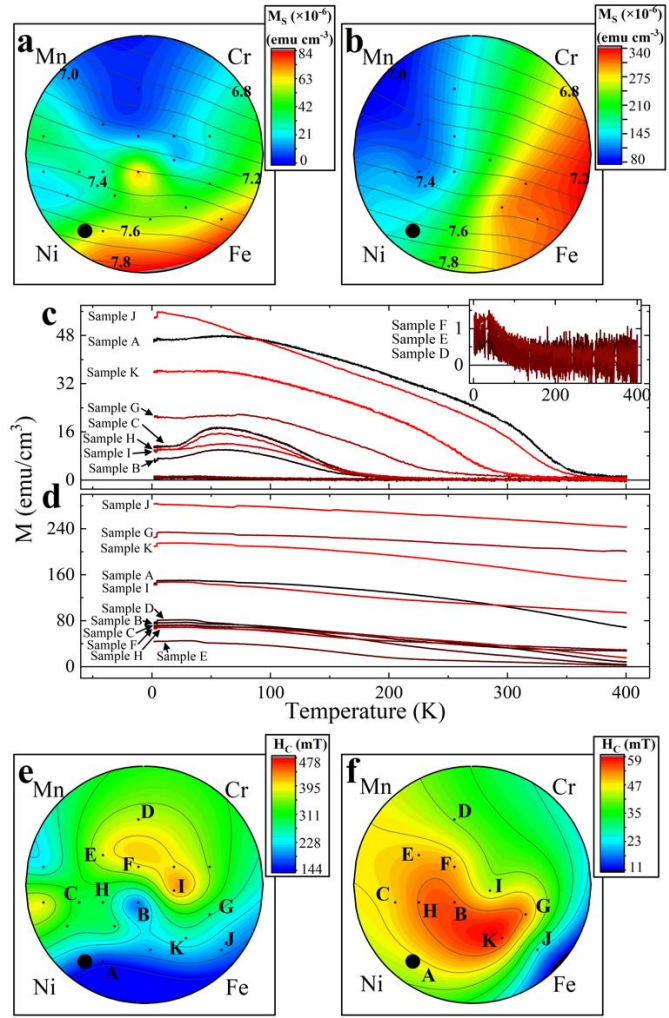

***Figure 4** Saturation magnetization, M-T plots and major loop coercivity at 2.5 K for (a, c, e) the as-grown sample and (b, d, f) the annealed sample. Contour lines in (a, b) indicate the average number of valence electrons, based on the composition.*

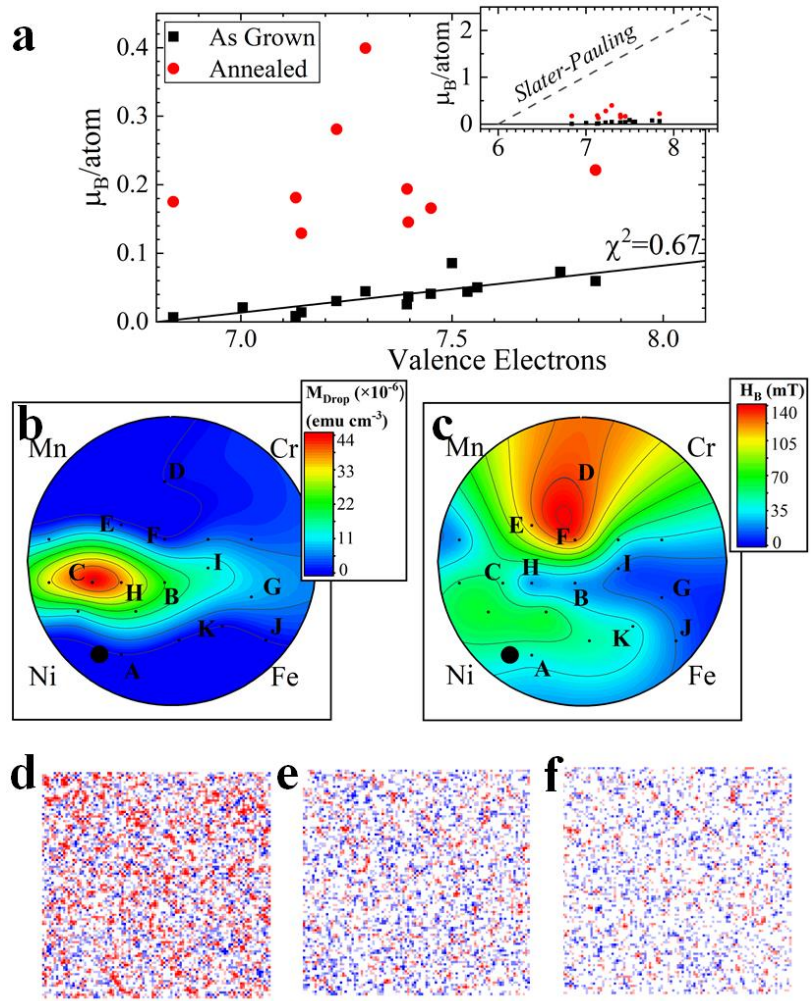

***Figure 5*** *(a) Saturation magnetization plotted against valence, with a comparison to the Slater-Pauling curve shown in the inset. The linear line fitted to the as-grown data has a reduced chi-squared value of 0.67. Heat maps of (b) the drop in the M-T plots and (c) the major loop exchange bias at 2.5 K. Simulated magnetic ordering in (d) Sample A, (e) Sample C and (f) Sample E, with red representing ferromagnetic clusters and blue antiferromagnetic.*

*Figure 6*

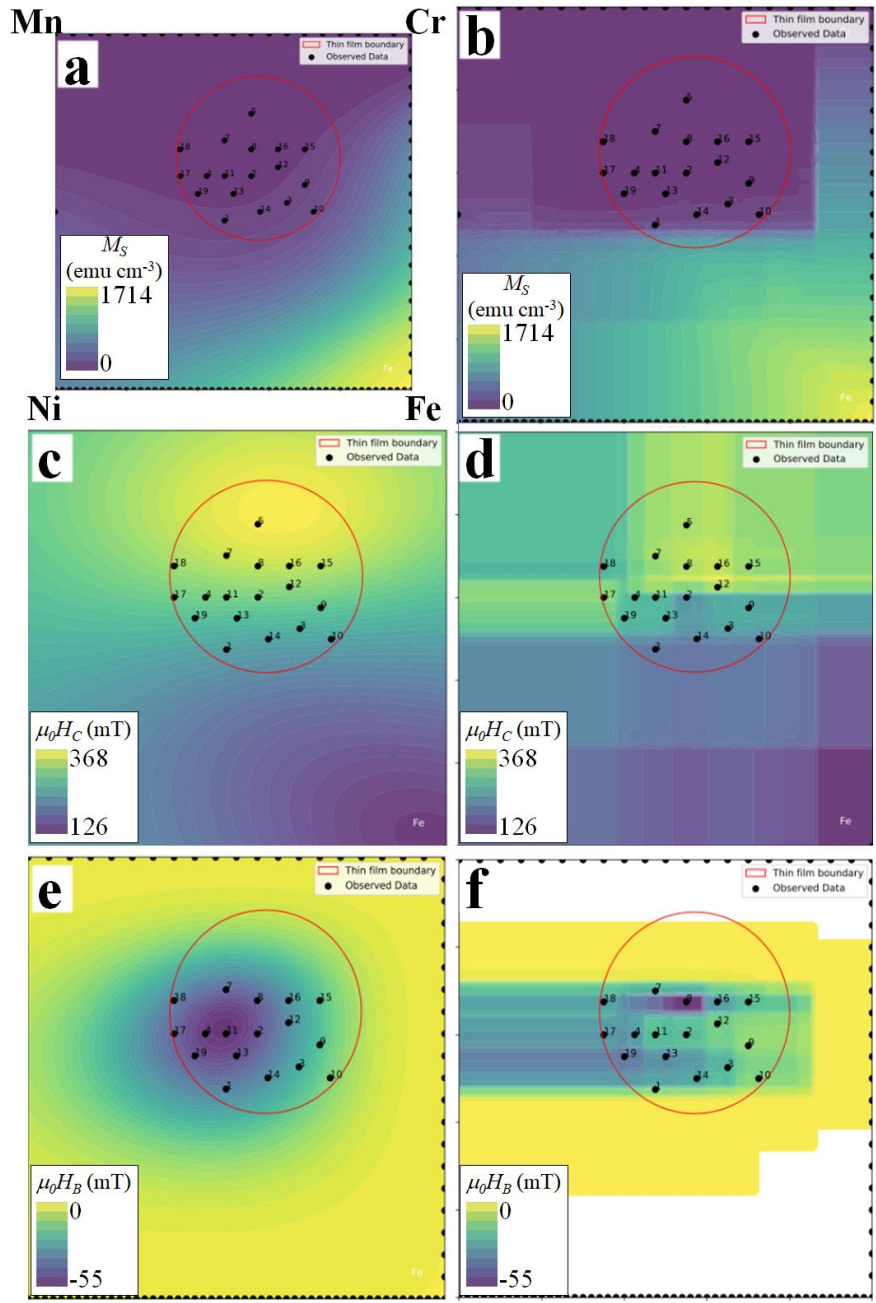

***Figure 6*** *Predicted values of (a, b) saturation magnetism, (c, d) coercivity, and (e, f) bias field across the (Cr, Mn, Fe, Ni) parameter space using Gaussian process regression (left) and random forest regression (right). The red circles denote the approximate boundary of the thin film wafer overlaid onto the parameter space. The black points denote the eighteen samples used to train the regression model (numbered 1-19, sample 6 was not measured). The contours show level sets of the predicted values. The black points on the perimeter of the parameter space represent theoretical values added to the training data.*